\documentstyle[preprint,prl,aps]{revtex}
\tightenlines
\begin{document}

\begin{center}
{{\large \bf Shape of nanosize superconducting grains: Does it influence pairing characteristics?
}}

\bigskip

V. N. Gladilin$^{1}$, V. M. Fomin$^{1}$, J. T. Devreese\footnote
{Corresponding author. E-mail: devreese@uia.ua.ac.be, fax: +32-3-8202245.

$^1$ 
Permanent address: Department of Theoretical Physics,
State University of Moldova,
Strada A.~Mateevici 60,
MD-2009 Kishinev, Moldova.

$^2$ 
Also at: Universiteit Antwerpen (RUCA), Groenenborgerlaan 171,
B-2020 Antwerpen, Belgium and
Technische Universiteit Eindhoven, P. O. Box 513,
5600 MB Eindhoven, The Netherlands.}$^{,2}$

\medskip

{\it TFVS, Departement Natuurkunde, Universiteit Antwerpen (UIA), 
Universiteitsplein 1, B-2610 Antwerpen, Belgium}

\medskip
\medskip

\medskip

{\bf Abstract}

\end{center}

\noindent
The Richardson exact solution for the reduced BCS Hamiltonian is applied to examine how sensitive are the pairing characteristics (condensation energy, spectroscopic gap, parity gap) to a specific configuration of single-electron energy levels in nanosize metallic grains. Using single-electron energy spectra in parallelepiped-shaped potential boxes with various volumes and aspect ratios as a model of energy levels in grains, we show that this sensitivity is extremely high. Just due to such an extreme sensitivity, the detailed shape of grains cannot be detected through the pairing characteristics, averaged over an ensemble of grains, even in the case of relatively small size dispersion within this ensemble. We analyse the effect of the pairing interaction on the excited-level spacings in superconducting grains and comment on the influence of shape-dependent fluctuations in single-electron energy spectra on the possibility to reveal this effect through tunnelling measurements.

\bigskip
\bigskip

\section{Introduction}

The experiments of Ralph, Black and Tinkham~\cite{brt1,brt2,brt3}, revealing, in particular, the presence of a parity-dependent spectroscopic gap in tunnelling spectra of nanometre-scale Al grains, inspired intensive theoretical studies of pairing correlations in those grains. 
Recently, the exact solution to the reduced BCS Hamiltonian, which was developed by Richardson in the context of nuclear physics as long ago as in 1963~\cite{RWR1}, has been reintroduced to the condensed matter community and used to describe superconductivity in nanosize metallic grains by Brown and von Delft~\cite{bd99a}. It is noteworthy that the Richardson solution is applicable at arbitrary distributions of single-electron energy levels, which form, together with the interaction strength, a complete set of input parameters for the reduced BCS Hamiltonian. This allows one to go beyond the model with equally spaced energy levels that is often used for nanosize superconducting grains. Thus, the Richardson solution has been applied~\cite{SDD00} to study the effect of level statistics on superconductivity in nanosize grains with random levels, assumed to follow the Gaussian orthogonal ensemble distribution, and to check the quality of the previous treatment of this effect~\cite{sa96}. Keeping in mind that an appropriate choice of statistics to describe the energy-level distribution in concrete nanostructures might be not self-evident, it seems interesting to analyse in more detail how sensitive are the pairing characteristics to a specific configuration of ``particle-in-a-box-like''~\cite{jvd96} single-electron states in grain. Based on the Richardson solution~\cite{RWR1}, in the present communication we tackle this problem by modelling single-electron energy spectra in grains with those spectra in parallelepiped-shaped potential boxes of various shape and size. The influence of the grain shape on the pairing characteristics (condensation energy, spectroscopic gap, parity gap) is studied both for single grains and for their ensembles. We also consider the effect of the pairing interaction on the excited-level spacings in superconducting grains and touch upon the influence of shape-dependent fluctuations of single-electron energy levels on the possibility to detect this effect through tunnelling measurements.

\section{The model}

For a nanosize grain, the BCS pairing Hamiltonian can be written as (see, e.~g., \cite{jvd96,ml97}) 
\begin{eqnarray}
H = \sum_{j,\sigma}\varepsilon_j a^{\dag}_{j\sigma}a_{j\sigma} -
\lambda d\sum_{i,j\in I } a^{\dag}_{i\uparrow}a^{\dag}_{i\downarrow}a_{j\downarrow}
a_{j\uparrow},
\label{HAM}\end{eqnarray}
where the operator $ a^{\dag}_{j\sigma}$ ($a^{ }_{j\sigma}$) creates (annihilates) an electron in the $j$-th time-reversed single-electron state with the spin $\sigma$ and the energy $\varepsilon_j$. The second term in Eq.~(\ref{HAM}) is the interaction Hamiltonian. The sum in this term is over the set $I$ of $J_I$ states inside the energy interval $(\varepsilon_{\rm F}- \hbar\omega_D, \varepsilon_{\rm F}+ \hbar\omega_D$), which will be referred to as the interaction band, with the Debye energy $\hbar\omega_D =34$~meV for Al~\cite{poole} and $\varepsilon_{\rm F}$, the Fermi energy. The interaction strength is a product of the mean energy-level spacing within the interaction band, $d=2\hbar\omega_D /J_I$, and the dimensionless parameter $\lambda$, taken to be 0.22 (close to that for Al \cite{bd98}). 

Since the electrons occupying levels outside the interaction band are straightforwardly described by the first term in the r.h.s. of Eq.~(\ref{HAM}), we will focus only on the electrons, which reside in the interaction band. Let the interaction band be populated by $N=2n+b$ electrons, of which $b$ electrons are on singly occupied levels (the set of these levels, blocked to the pair scattering, is denoted as $B$). The electrons on singly occupied levels do not participate in the pair scattering and their contribution to the energy of the $N$-electron system under consideration is simply $\sum_{j\in B} \varepsilon_j$. The remaining $2n$ electrons form $n$ pairs, which are distributed among the set $U= I\setminus B$ of $J_I-b$ unblocked levels. Richardson~\cite{RWR1} showed that the energy of these pairs is given by the sum of $n$ parameters $E_{1}, \ldots, E_{n}$, which are non-degenerate roots ($E_{\mu}\neq E_{\nu}$ for all $\mu\neq\nu$) of the following set of $n$ coupled equations:
\begin{eqnarray}
\frac{1}{\lambda}-\sum_{j\in U}\frac{d}{2\varepsilon_j - E_{\nu}} +
\sum_{\mu=1\atop\mu \neq \nu}^{n}\frac{2d}{E_{\mu}- E_{\nu}} =0, \\
\nu=1,\ldots ,n\;.\nonumber
\label{R1}\end{eqnarray}
Thus the total energy of electrons, which populate the interaction band, is
\begin{eqnarray}
E_{N,b}= \sum_{j\in B} \varepsilon_j +\sum_{\nu=1}^n E_{\nu}, \qquad N=2n+b. 
\label{etot}\end{eqnarray}

Given the energies $E_{N,b}$, the {\it pairing characteristics} like the condensation energy, the spectroscopic gap, and the parity gap can be calculated. Our first aim is to analyse how sensitive are these characteristics to a specific configuration of single-electron energy levels, $\varepsilon_j$, in the interaction band of a grain. As an example, we will consider the single-electron energy spectra in a parallelepiped-shaped hard-wall box with sizes $l_x$, $l_y$ and $l_z$:   
\begin{eqnarray}
\varepsilon_j=\frac{\pi^2\hbar^2}{2m}\left(\frac{ n_x^2(j)}{l_x^2} +\frac{ n_y^2(j)}{l_y^2} +\frac{ n_z^2(j)}{l_z^2} \right), \label{eps}
\\
n_x(j),n_y(j),n_z(j)=1,2,3,...,\nonumber
\end{eqnarray}
where $m$ is the effective mass of an electron ($m=1.4m_e $ in Al \cite{poole}). The correspondence between $j$ and  $n_x(j)$, $n_y(j)$, $n_z(j)$ is determined by the requirement $\varepsilon_j \leq\varepsilon_{j+1}$, which we impose. When labelling the single-electron energy levels, we will take $j=0$ for the uppermost energy level populated by electrons in the ground state.  A grain with sizes $l_x$, $l_y$ and $l_z$ contains 
$N_{\rm tot}=n_el_xl_yl_z $ electrons, where the electron density is taken to be $n_e=181$ nm$^{-3}$ for Al~\cite{poole}. A major part of these electrons are on energy levels below the interaction band,
which are assumed to be completely filled. Assuming in addition that all the energy levels with $\varepsilon_j > \varepsilon_{\rm F}+ \hbar\omega_D$ are empty, the number of electrons in the interaction band, $N$, is fully determined by the volume and the shape of the grain, the parity of $N$ being the same as that of $ N_{\rm tot}$.

\section{Shape-dependent pairing characteristics of a nanosize grain}

First, we will consider the condensation energy, which can be expressed as (see, e.~g., Refs.~\cite{SDD00,schech})    
\begin{eqnarray}
E^{\rm C}_{N,b}(\lambda)= E^{\rm G}_{N,b}(0)- E^{\rm G}_{N,b}(\lambda)-n\lambda d, 
\label{eint}\end{eqnarray}
where $E^{\rm G}_{N,b}$ is the ground-state energy of the $N$-electron system in the interaction band ($b=0$ for even $N=2n$ and $b=1$ for odd $N=2n+1$). 
In Fig.~1 we show the calculated condensation energy for nearly cubic grains with a fixed number of electrons, $N_{\rm tot}=4000$ (the corresponding volume of grains is approximately 22~nm$^{3}$), when varying the aspect ratio, $ l_x: l_y: l_z =1: (1+\eta): (1+2\eta)$. 
As seen from Fig.~1, the condensation energy exhibits well-pronounced maximums at some values of the parameter $\eta$. An inspection of the corresponding single-electron energy spectra, $\varepsilon_j$, shows that the maximums appear when a degeneracy or quasi-degeneracy occurs between the lowest empty energy levels and the uppermost occupied energy levels. The degree of degeneracy is especially strong in a perfect cube resulting in a huge (about 104 meV high) peak of the condensation energy at $\eta=0$. However, this peak is seen to be extremely narrow implying that a proposal~\cite{bgk00} for an experimental observation of degeneracy effects typical for perfectly symmetric grains can be hardly realised. At the same time, even for less symmetric grains, the variations of the condensation energy as a function of the grain shape are still strong (about one order of magnitude) reflecting a high sensitivity of the pairing characteristics to the detailed structure of the single-electron energy spectrum in the vicinity of the Fermi level. As illustrated by Fig.~2, this sensitivity is also manifested when considering the condensation energy for grains with a fixed aspect ratio as a function of their volume or, equivalently, of $N_{\rm tot}$. 

Figure~2 shows also the lowest excitation energies,
\begin{eqnarray}
d^{(i)}_{N}=\left \{  { 
E^{(i)}_{N,1}-E^{\rm G}_{N,1} \ {\rm for}\  N =2n+1, \atop 
E^{(i)}_{N,2}-E^{\rm G}_{N,0} \ \ \ \ \ \ {\rm for}\ N =2n,} \right. \label{eexc}\\
i=1,2,3,\ldots, \nonumber
\end{eqnarray}
relevant to the so-called ``primary'' peaks \cite{schech} in tunnelling spectra of nanosize superconducting grains. In grains with an odd number of electrons, the energies $E^{(i)}_{N,1}$ correspond to excited states with a single unpaired electron on the $i$-th level ($i>0$). For a grain with an even number of electrons, the energies $E^{(i)}_{N,2}$ correspond to excited states, where two electrons are unpaired, one of them having the energy $\varepsilon_0$ and another occupying the $i$-th single-electron energy level. Since these excited states involve one broken Cooper pair, the differences between their energies and the energy of the fully paired ground state are at least as large as the energy cost for pair breaking. The quantity $d^{(1)}_{N}$ defined for an even $N$ coincides with the spectroscopic gap~\cite{bd99}. Figure~2 allows one to see a close similarity between the behaviour of the condensation energy as a function of $N_{\rm tot}$ and that of the contribution to the spectroscopic gap due to the pairing interaction. Another observation, implied from Fig.~2, is an increase of the density of excited energy levels in a superconducting grain as compared to that in a normal grain (with $\lambda=0$).  
The physical reason for this increase is the blocking of energy levels by unpaired electrons that reduces the interaction energy of the remaining electron pairs. With increasing $i$, the effect of the $i$-th level blocking on the interaction energy relaxes. Therefore, a downward energy shift due to the pairing interaction is larger for higher excited states than that for lower ones. This means that the energy spacing between excited states decreases when the interaction is switched on. Such a decrease was recently analysed within a perturbative approach for highly excited states ($\varepsilon_i-\varepsilon_{0}\gg d$) in the case of equidistant single-electron energy levels~\cite{schech}. However, it follows from the above discussion as well as from Fig.~2 that an increase of the energy-level density due to the pairing interaction is more pronounced just for the lowest excited states and for small inter-level spacings $\varepsilon_i-\varepsilon_{0}$, when a perturbative treatment is hardly applicable. 

In Fig.~3, we plot the calculated parity gap~\cite{ml97}, 
\begin{eqnarray}
\Delta^{\rm P}_n= E^{\rm G}_{2n+1,1}-\frac12\left(E^{\rm G}_{2n,0} + E^{\rm G}_{2n+2,0} \right), 
\label{pgap}\end{eqnarray}
i.~e. the difference between the ground-state energy of a grain, which contains an odd total number of electrons, $2n+1$ of them being in the interaction band, and the mean ground-state energy of the grains obtained by removing or adding one electron. Since Eq.~(\ref{pgap}) involves the ground-state energies of three systems with different number of electrons, the behaviour of $\Delta^{\rm P}_n$ as a function of $N_{\rm tot}$ is somewhat richer as compared to that of the condensation energy, $E^{\rm C}_{2n+1,1}$, which is also shown in Fig.~3.

\section{Pairing characteristics for ensembles of nanosize grains}

Having demonstrated an extreme sensitivity of pairing characteristics 
to the specific geometry of a {\it single} grain, we will now analyse these characteristics for {\it ensembles} of nanosize grains.  Let us consider a set of parallelepiped-shaped grains whose sizes are given by expressions
\begin{eqnarray}
l_{\alpha}=\langle l_{\alpha} \rangle+r_{\alpha} \Delta l, \quad {\alpha}=x,y,z
\end{eqnarray}
with random values $r_{\alpha} \in (-1, 1)$. Since we will deal with values $\Delta l\le 0.25$ nm, which are expected to be much smaller than the size dispersion for ensembles of realistic grains, all values of $r_{\alpha}$ within the interval ($-1, 1$) can be taken equally probable.  
For several ensembles of 500 to 700 grains with $\langle l_x \rangle: \langle l_y \rangle:\langle l_z \rangle=1:2:2$ and $\Delta l=0.2$ nm, the ensemble average for the condensation energy, $\langle E^{\rm C}_{N,0} \rangle$, and the variance $\delta E^{\rm C}_{N,0}\equiv \left[\langle (E^{\rm C}_{N,0})^2 \rangle -\langle E^{\rm C}_{N,0} \rangle^2 \right]^{1/2} $ are shown in Fig.~4(a) as a function of $\langle l_x \rangle$. Analogous plots for the spectroscopic gap and its variance are given in Fig.~4(b). For comparison, we display the condensation energy and the spectroscopic gap in grains, where single-electron energy levels in the interaction band are equally spaced by $\langle d \rangle$, the average energy distance between adjacent levels for corresponding ensembles. 
For grains with equidistant levels $\varepsilon_j$, the condensation energy is appreciably smaller than $\langle E^{\rm C}_{N,0} \rangle$.  
As concerns the spectroscopic gap, the difference between the results for grains with equidistant levels and those for ensembles of grains is less pronounced. There are two (positive) contributions to the spectroscopic gap: one is due to the energy spacing $\varepsilon_1-\varepsilon_0$, another is determined by the coupling energy of an electron pair. Since the coupling energy increases with decreasing $\varepsilon_1-\varepsilon_0$, fluctuations of these two contributions within an ensemble of grains partially cancel each other. 

It is worth noting that for the ensembles of grains under consideration, the variances $\delta E^{\rm C}_{N,0}$ and $\delta d^{(1)}_{N}$ as well as the differences between the ensemble averages $\langle E^{\rm C}_{N,0} \rangle$ ($\langle d^{(1)}_{N} \rangle$) and the corresponding values $E^{\rm C}_{N,0}$ ($d^{(1)}_{N}$) in grains with equidistant levels are several times larger as compared to the results of Ref.~\cite{SDD00}, where single-electron energy levels in grains were assumed to follow the Gaussian orthogonal ensemble (GOE) distribution~\cite{rm}. The reason for this discrepancy is that for the single-electron energy spectra given by Eq.~(\ref{eps}), energy-level spacings obey the Poisson distribution. In particular, small nearest-neighbour spacings ($\varepsilon_j-\varepsilon_{j-1}\ll d$), which are unfeasible in GOE, have the maximum probability density in the ensembles analysed here.  It is interesting to address the question of whether or not the ``average grain shape'', which is given in our example by the aspect ratio $\langle l_x \rangle: \langle l_y \rangle:\langle l_z \rangle$, can be revealed through the pairing characteristics of an ensemble of grains. Since the variances $\delta E^{\rm C}_{N,0}$ and $\delta d^{(1)}_{N}$ are relatively large, the answer appears to be negative. Thus, the relative differences between the averages, calculated for three ensembles of grains with approximately one and the same volume ($\langle l_x \rangle \langle l_y \rangle\langle l_z \rangle=172.5$~nm$^3$, $\Delta l=0.2$~nm) but substantially different aspect ratios ($\langle l_x \rangle: \langle l_y \rangle:\langle l_z \rangle=1:1:1$,  $1:2:2$, and $1:2:16$), do not exceed 2.5\% for  $\langle E^{\rm C}_{N,0} \rangle$ and 3\% for $\langle d^{(1)}_{N} \rangle$.  

Let us analyse the aforementioned effect of the pairing interaction on the spacings between excited states with energies $E^{(i)}_{N,b}$ ($b=1$ for odd $N$ and $b=2$ for even $N$). Counting the energy from the lowest excited level, $E^{(1)}_{N,b}$, we can write down the density of higher energy levels as
\begin{eqnarray}
D(E)=\sum\limits_{i\geq 2}\delta\left(E-\left[E^{(i)}_{N,b}- E^{(1)}_{N,b}\right]\right).
\label{dens}\end{eqnarray}
Since the energy spectra are strongly affected even by weak fluctuations of the grain shape, modifications of the ``detailed'' level density $D(E)$ due to the pairing interaction might be hidden by seemingly irregular distribution of energy levels, especially when studying a small grain with only a few energy levels in the interaction band. As a more convenient quantity, we introduce the excited-level density averaged over a relatively large energy interval $\Delta$:  
\begin{eqnarray}
D_\Delta(E)\equiv \frac{1}{\Delta}
\int\limits_{[E/\Delta]\Delta}^{([E/\Delta]+1)\Delta }D(E)\,dE, 
\label{avdens}\end{eqnarray}
where $[x]$ denotes the integer part of $x$. To reveal more clearly the influence of the pairing interaction on the excited-level density, we examine the function
\begin{eqnarray}
\rho_{\Delta,{\cal E}}(E)= \frac{ D_\Delta(E)- D_{\cal E }(0)}{ D_{\cal E }(0)},
\label{rho}\end{eqnarray}
which describes relative deviations of $ D_\Delta(E)$ from the level density averaged over an energy interval ${\cal E}\gg \Delta$. 
In Fig.~5, we show the ensemble averages, $\langle \rho_{\Delta,{\cal E}}\rangle$, calculated for two ensembles of 1000 nanosize grains ($\langle l_x \rangle \langle l_y \rangle\langle l_z \rangle=23.3$~nm$^3$) with an even [panel(a)] and odd [panel(b)] number of electrons. In spite of fluctuations of $\langle \rho_{\Delta,{\cal E}}\rangle$, which are still present even for relatively large ensembles under consideration, a pronounced trend to an increase of the excited-level density when lowering $E$ is seen from plots for superconducting ($\lambda\neq 0$) grains at any parity of the number of electrons. In this respect, the behaviour of $\langle \rho_{\Delta,{\cal E}}\rangle$ for superconducting grains is quite distinctive from that for corresponding normal ($\lambda= 0$) grains. The inset in Fig.~5(b) demonstrates that for largish grains (with $\langle l_x \rangle \langle l_y \rangle\langle l_z \rangle=256$~nm$^3$) the aforementioned distinction can be revealed even when averaging the function $ \rho_{\Delta,{\cal E}} $ over a few-grain set. The obtained results imply that the excited-level densening due to the pairing interaction might be detected through experimental tunnelling spectra accumulated over sets of grains.      

\bf{Acknowledgements}\rm{ -}
This work has been supported by the GOA BOF UA 2000, I.U.A.P., F.W.O.-V.
projects Nos. G.0306.00, G.0274.01N and the W.O.G. WO.025.99N
(Belgium).


\vspace{0.5 truecm}

\begin{center}
  {\bf Figure captions}
\end{center}

\vspace{0.5 truecm}

\noindent Fig. 1. Condensation energy $E^{\rm C}_{N,0}$, calculated according to Eq.~(\ref{eint}) for paral\-le\-le\-piped-shaped grains with the aspect ratio $ l_x: l_y: l_z =1: (1+\eta): (1+2\eta)$, as a function of the parameter $\eta$. The volume of grains is fixed to keep a constant number of electrons in a grain, $N_{\rm tot}=4000$. The number of electrons in the interaction band, $N$, varies from 20 to 42 for the range of $\eta$ under consideration.

\vspace{0.2 truecm}

\noindent Fig. 2. Condensation energy $E^{\rm C}_{N,0}$, given by Eq.~(\ref{eint}), and the lowest excitation energies $d^{(i)}_{N}$ ($i=1$ to 5), given by Eq.~(\ref{eexc}), for parallelepiped-shaped grains with the aspect ratio $l_x: l_y: l_z =1:1.3:1.7$ as a function of the number of electrons $N_{\rm tot}$. Lines are guides for eye. Only grains with even $N_{\rm tot}$ are considered. The number of electrons in the interaction band, $N$, varies from 108 to 146 depending on $N_{\rm tot}$. For comparison, the five lowest excitation energies for the same grains in their normal state (at $\lambda=0$) are also shown. 

\vspace{0.2 truecm}

\noindent Fig. 3. Parity gap $\Delta^{\rm P}_n$, given by Eq.~(\ref{pgap}),  and condensation energy $E^{\rm C}_{2n+1,1}$, given by Eq.~(\ref{eint}), for parallelepiped-shaped grains with the aspect ratio $l_x: l_y: l_z =1:1.9:2.1$ as a function of the (odd) number of electrons $N_{\rm tot}$. Lines are guides for eye. The number of pairs in the interaction band, $n$, varies from 83 to 110 for grains under consideration.

\vspace{0.2 truecm}

\noindent Fig. 4. Ensemble averages and variances of the condensation energy [panel (a)] and of the spectroscopic gap [panel (b)] as a function of the average grain size $\langle l_x \rangle$ for ensembles of 500 to 700 grains with $\langle l_x \rangle: \langle l_y \rangle:\langle l_z \rangle=1:2:2$ and $\Delta l=0.2$~nm. The condensation energy and the spectroscopic gap are displayed also for grains with equidistant single-electron energy levels in the interaction band, the spacing between these levels coinciding with the average value $\langle d \rangle$ for the corresponding ensemble. The ensemble average for the number of electrons in the interaction band, $\langle N \rangle $, varies from 16 at $\langle l_x \rangle=1.5$ nm to 714 at $\langle l_x \rangle=5.5$ nm. 

\vspace{0.2 truecm}

\noindent Fig. 5. Ensemble average $\langle \rho_{\Delta,{\cal E}}\rangle$,   where $\rho_{\Delta,{\cal E}}$ is given by Eq.~(\ref{rho}), for 1000 superconducting grains ($\lambda=0.22$) with $\langle l_x \rangle=1.8$~nm, $\langle l_y \rangle=3.5$~nm, $\langle l_y \rangle=3.7$~nm, and $\Delta l=0.25$~nm is shown as a function of $E$ in the case when the number of electrons in each grain is even [panel(a)] and odd  [panel(b)]. Inset: function $\rho_{\Delta,{\cal E}}(E)$, averaged over a set of 20 grains with $\langle l_x \rangle=4$~nm, $\langle l_y \rangle=8$~nm, $\langle l_y \rangle=8$~nm, and $\Delta l=0.2$~nm; the number of electrons in each grain is odd. 
For comparison, the functions $\langle \rho_{\Delta,{\cal E}}(E)\rangle$, calculated for corresponding normal grains ($\lambda=0$) are also shown.

\end{document}